\documentclass[aps,prb,twocolumn,superscriptaddress,floatfix,a4]{revtex4}
\usepackage{epsfig,amsmath,amssymb,color}
\begin{document}

\title{A real-time   study of diffusive and ballistic
transport in spin-$1/2$ chains using the adaptive time-dependent density matrix renormalization group method}
\author{S. Langer}
\affiliation{Institut f\"ur Theoretische Physik C, RWTH Aachen University, 52056 Aachen, Germany, and 
J\"ulich Aachen Research Alliance (JARA), Research Centre J\"ulich GmbH, 52425 J\"ulich, Germany}
\author{F. Heidrich-Meisner} 
\email[Corresponding author: ]{fabian.heidrich-meisner@physik.rwth-aachen.de}
\affiliation{Institut f\"ur Theoretische Physik C, RWTH Aachen University, 52056 Aachen, Germany, and
JARA -- J\"ulich Aachen Research Alliance, Forschungszentrum J\"ulich, Germany}
\author{J. Gemmer} 
\affiliation{Institut f\"ur Theoretische Physik, Universit\"at Osnabr\"uck, Germany}
\author{I.P. McCulloch} 
\affiliation{School of Physical Sciences, The University of Queensland, Brisbane, Queensland 4072, Australia }
\author{U. Schollw\"ock} 
\affiliation{Physics Department and Arnold Sommerfeld Center for Theoretical 
Physics, Ludwig-Maximilians-Universit\"at M\"unchen, D-80333 M\"unchen, 
Germany}

\date{June 16, 2009}

\begin{abstract}
Using the adaptive time-dependent density matrix renormalization group method, 
we numerically study the spin dynamics and transport in one-dimensional spin-1/2 systems at zero temperature. 
Instead of computing transport coefficients
from linear response theory, we study the real-time evolution of the magnetization  starting from
spatially inhomogeneous initial states. In particular, we are able to analyze systems far away from equilibrium with this set-up. By computing the time-dependence of the variance of 
the magnetization, we can distinguish diffusive from ballistic regimes, depending on model parameters.
For the example of the anisotropic spin-1/2 chain  and at half filling, we find the expected ballistic behavior in the
easy-plane phase, while in the massive regime the dynamics of the magnetization  is   diffusive.
Our approach allows us to tune the deviation of the initial state from the ground state and the qualitative behavior of the dynamics turns out to be valid even
for highly perturbed initial states in the case of easy-plane exchange anisotropies.
We further cover two examples of nonintegrable models, the frustrated chain and the two-leg spin ladder, and we encounter diffusive 
transport in all massive phases. In the former system, our results indicate ballistic behavior in the critical phase.
We propose that 
the study of the time-dependence of the spatial variance of particle densities could be instrumental in the 
characterization of the expansion of ultracold atoms in optical lattices as well.
\end{abstract}

\maketitle
\section{Introduction}
Transport in low-dimensional strongly correlated systems continues to excite 
theoretical and experimental physicists alike. For theorists, transport problems pose 
a formidable challenge, as  established tools to work out ground-state properties of
strongly correlated systems do not always provide an adequate description of transport as well (see Refs.~\onlinecite{zotos-review}
and \onlinecite{hm07a} and references therein), which, in particular, pertains to systems driven out of equilibrium.

Within linear-response theory, one often distinguishes between ballistic and diffusive transport
by invoking the notion of the  Drude weight,\cite{kohn64,scalapino92} {\it i.e.}, the prefactor of a delta-function at zero-frequency 
in the frequency dependent transport coefficient. A finite Drude weight defines ballistic transport, while
in the case of a vanishing Drude weight, the zero-frequency limit of the conductivity's regular part 
determines the long-time behavior.

Significant theoretical attention has been devoted  to one-dimensional spin systems (see
Refs.~\onlinecite{zotos-review,zotos05,hm07a} for a review). 
Open theoretical questions include, for instance, 
the finite temperature transport of the anisotropic spin-1/2 chain with nearest-neighbor interactions (the 
$XXZ$ chain), 
with an unsettled debate  on whether finite temperature
transport in the Heisenberg chain is ballistic or not \cite{narozhny98,fabricius98,zotos99,alvarez02,hm03,rabson04,benz05,heidarian07}  
as well as on the actual temperature dependence of the Drude weight.\cite{benz05,zotos99,fujimoto03}
For nonintegrable models and finite temperatures, one expects diffusive transport on general grounds,\cite{castella95,rosch00,hm07a,zotos05} 
and numerical studies have widely confirmed this picture in the high-temperature limit and massive phases
of spin models.\cite{zotos96,narozhny98,hm03,rabson04,mukerjee06,mukerjee08} A similar scenario has emerged 
for thermal transport.\cite{hm02,hm03,hm04,zotos04}
Yet,  the issue of (quasi)ballistic transport 
in gapless phases of, {\it e.g.}, the frustrated chain,\cite{jung06,heidarian07,hm03,jung07a} 
at low temperatures is still under scrutiny. Moreover, the possibility of
  anomalous transport due to a diverging 
coefficient of the dc conductivity has been emphasized.\cite{zotos04,prelovsek04,jung06}

Much less is known about the non-linear transport at large external driving forces, or more generally,
nonequilibrium properties. A recent study using a quantum master-equation approach has addressed the spin transport
in the antiferromagnetic phase of the $XXZ$ chain.\cite{benenti09} 
The time-evolution of magnetization profiles in analytically exactly solvable models has been the case of
interest 
 in Refs.~\onlinecite{antal99,berim02,ogata02,hunyadi04}.

Besides the fundamental interest in understanding 
large-bias and out-of-equilibrium phenomena, research into transport properties of low-dimensional spin systems 
is strongly motivated by exciting experimental results on large thermal conductivities in spin ladder
and chain materials (see, e.g., Refs.~\onlinecite{sologubenko07a,hess07a} for a review). Most evidently in spin ladder materials
such as (La,Sr,Ca)$_{14}$Cu$_{24}$O$_{41}$, such large thermal conductivities have been attributed to
magnetic excitations.\cite{sologubenko00,hess01} 
In a more recent experiment on La$_5$Ca$_{9}$Cu$_{24}$O$_{41}$,\cite{otter09}
heat dynamics has been probed by time-of-flight measurements. In this set-up, the surface of a sample is covered with a thin fluorescent layer.
After shining on that surface with a laser, one can then follow the propagation of heat in the surface by thermal imaging at different times.
A pronounced difference is seen comparing a surface that contains ladders to one that is perpendicular to the ladder direction.
In the former case, heat diffuses predominantly along the ladder direction, while little dynamics is seen in the latter case.
These results support the notion of anisotropic heat transport in this material,\cite{hess01} due to the 
contribution of magnetic excitations.

Besides thermal transport measurements, 
questions of diffusive versus ballistic transport have  been experimentally probed in nuclear magnetic
resonance\cite{takigawa96, thurber01} (see Ref.~\onlinecite{sirker06} for related theoretical work) as 
well as muon spin resonance experiments.\cite{pratt06} In a more recent development, transport properties of low-dimensional ultra-cold atom gases have gained 
attention as well, with experiments focusing on the detection of Anderson localization.\cite{fort05,clement05}
Interacting two-component Bose gases in optical lattices have been suggested to potentially
realize spin-1/2 Hamiltonians.\cite{trotzky08,barthel08a}

As far as numerical approaches are concerned, on the technical side, full exact diagonalization (ED) studies are restricted to system sizes of about 24 sites in the case of
a spin-1/2 chain, while in Quantum Monte Carlo simulations, the calculation of frequency dependent 
properties of more than two-point
correlation functions remains difficult (see, {\it e.g.}, Ref.~\onlinecite{grossjohann08} and references therein). The density matrix renormalization group (DMRG)
method\cite{white92b,white93,schollwoeck05}
has most successfully been applied to zero-temperature phenomena. With the advent of the adaptive time-dependent 
DMRG method (tDMRG),\cite{daley04,white04,vidal04} the study of non-equilibrium and large bias phenomena
has become possible. While the methods mentioned so far are designed for pure bulk systems, 
 quantum-master approaches that account for dissipation by
incorporating baths have been used as well in the study of transport in 1D spin 
systems.\cite{saito02,steinigeweg06,meija-monasterio07,michel08,prosen09} Such methods are not constrained to the linear
response either and circumvent the use of Kubo formulae, as they measure 
gradients and current expectation values directly.

 In this work we will introduce and exploit an alternative approach 
based on zero-temperature tDMRG calculations. Instead of analyzing currents and their correlation
functions directly,
we study the magnetization dynamics after preparing the system in inhomogeneous initial states.
For instance, we subject the system to an external magnetic field of
Gaussian shape and, after releasing the confining field, follow the 
time evolution of the magnetization. 
Computing the variance of the magnetization allows us to distinguish
ballistic from diffusive regimes, depending on model parameters: we 
 consider the dynamics to be ballistic if the variance grows quadratically in time,
which is the behavior of noninteracting particles, while a diffusive behavior
manifests itself in a linear increase of the magnetization's variance. 
Our approach has the advantage that we can characterize diffusion by following
the time-evolution of a local quantity, the magnetization,  as compared to the
technically more difficult  evaluation of the Kubo formula\cite{kuehner00} or the measurement of time-dependent currents.\cite{gobert05} Moreover, we can control the deviation of the initial state from the
ground state, thus scanning the  regime of systems substantially driven out of equilibrium. 

While we show that our results for the spin-1/2 chain with an exchange anisotropy
in the regime of small perturbations over the ground-state are consistent with the
picture established by analyzing Drude weights at zero temperature, namely
ballistic transport in the massless and diffusive transport in the massive regime,\cite{shastry90} 
we, in particular, argue that this also applies to systems  far from equilibrium.
The dynamics is further sensitive to the overall filling, or average magnetization, as expected from linear-response theory 
results for the high-temperature limit.\cite{zotos97,louis03,hm05,sakai05}

Beyond the anisotropic spin-1/2 chain with nearest-neighbor interactions only,
we further consider nonintegrable systems such as the frustrated chain and the 
two-leg spin  ladder. As a result, we find that in massive phases, the dynamics is
typically diffusive, while in the massless one of the frustrated chain, the zero temperature dynamics 
are ballistic. 

Transport in the $XXZ$ chain has previously been studied in Ref.~\onlinecite{gobert05} using tDMRG,
 there following the  time-evolution from a highly excited initial state
 of the  $|\psi \rangle=| \uparrow\dots \uparrow\downarrow \dots\downarrow\rangle $ form.
The long-time behavior of the magnetization was found to be correlated with the 
phase transition from easy-plane to easy-axis symmetry.
Further, 
the expansion of particle density packets of nearly Gaussian form has been looked at with tDMRG
in the context of ultra-cold atomic gases,\cite{kollath05b} modeled with the 1D Bose-Hubbard,
as well as for short pieces of interacting spinless fermions.\cite{schmitteckert04}
 
The plan of the paper is the following. In Sec.~\ref{sec:model} 
we define the spin models studied and we describe our numerical method,
the tDMRG.  We further motivate our definition of diffusive transport by 
discussing the solution of the diffusion equation  in Sec.~\ref{sec:theory}.  Section~\ref{sec:init} details the preparation of initial states.
In Sec.~\ref{sec:xxz}, we study the magnetization dynamics in the $XXZ$ chain,
 with ballistic transport in the massless regime, and diffusive transport in the massive regime.
Section~\ref{sec:non} summarizes our results for two nonintegrable 1D systems, 
the frustrated chain and the two-leg ladder. We conclude with a discussion in
Sec.~\ref{sec:sum}.

\section{Model and method}
\label{sec:model}
\subsection{1D Spin-$1/2$ systems}
Here we will first concern ourselves with the integrable $XXZ$ chain:
\begin{equation}
H = \sum_{i=1}^{L-1} \lbrack\frac{1}{2}(S_{i}^+S_{i+1}^{-}+H.c.) + \Delta S^z_iS^z_{i+1} \rbrack\, , \label{eq:xxz}
\end{equation}
where $S_i^{\mu}$ and$\mu=x,y,z$ are the components of a spin-$1/2$ operator acting on site $i$
and $S^{\pm}_i= S_i^{x}\pm i S^y_i$ are the lowering/raising operators, respectively.
We denote the number of sites by $L$ and we introduce an exchange anisotropy $\Delta$.
Equation \eqref{eq:xxz} can be re-expressed in terms of spinless fermions $c_i^{(\dagger)}$ through the 
Jordan-Wigner transformation:\cite{mahan}
\begin{equation}
H = \sum_{i=1}^{L-1} \lbrack\frac{1}{2}(c_{i}^{\dagger}c_{i+1}^{}+H.c.) + \Delta (n_i-1/2) (n_{i+1}-1/2) \rbrack\, , \label{eq:xxz_c}
\end{equation}
with $n_i=c_i^{\dagger}c_i$. Setting $\Delta=0$ results in a noninteracting system. If not mentioned otherwise, we 
impose open boundary conditions. We denote the filling factor with $n$. The local magnetization is given by $M_i=S^z_i$, and the total magnetization is
$S^z=\sum_i S_i^z = L(n-1/2)$.

The ground-state phase diagram of the $XXZ$ chain (see, {\it e.g.}, Ref.~\onlinecite{review-frust}
 and references therein)  exhibits quantum critical points at $\Delta=\pm 1$. A critical phase
covers the $|\Delta| \leq 1$ region, while the ground state for $\Delta>1$ exhibits antiferromagnetic
order. The region $\Delta\leq -1$ has a ferromagnetic ground-state, yet we will restrict the discussion to $\Delta \geq 0$. The model is integrable through the Bethe-ansatz.\cite{kluemper-review}

In the second part of this work, we will focus on two nonintegrable models with isotropic interactions (i.e., $\Delta=1$), the frustrated chain and the 
two-leg ladder (for a review on these models, see, {\it e.g.}, Refs.~\onlinecite{dagotto99,review-frust}). 
Both models can be understood as limiting cases of a single Hamiltonian that incorporates a dimerization $\delta$ and a frustration $\alpha$:
 \begin{equation}
H = J \sum_{i=1 \atop \mu=x,y,z}^{L-1} \lbrack (1+(-1)^{i}\delta) S_{i}^{\mu}S_{i+1}^{\mu} + \alpha
S_{i}^{\mu}S_{i+2}^{\mu}\rbrack \,.\label{eq:non}
\end{equation}
The frustrated chain corresponds to $\delta=0$ and $\alpha >0$, while the two-leg ladder is the $\delta=1$ limit.
In the latter case, we identify the coupling along legs as $J_{\parallel}=\alpha J$ 
and the coupling along rungs as $J_{\perp}=J$. 

The frustrated chain  features a quantum phase transition at $\alpha \approx 0.241$, separating a gapless phase from a massive one.\cite{review-frust}
The spectrum of the two-leg ladder is gapped for any $ J_{\perp}/J_{\parallel}> 0 $.\cite{dagotto96,dagotto99}

\begin{figure*}[t]
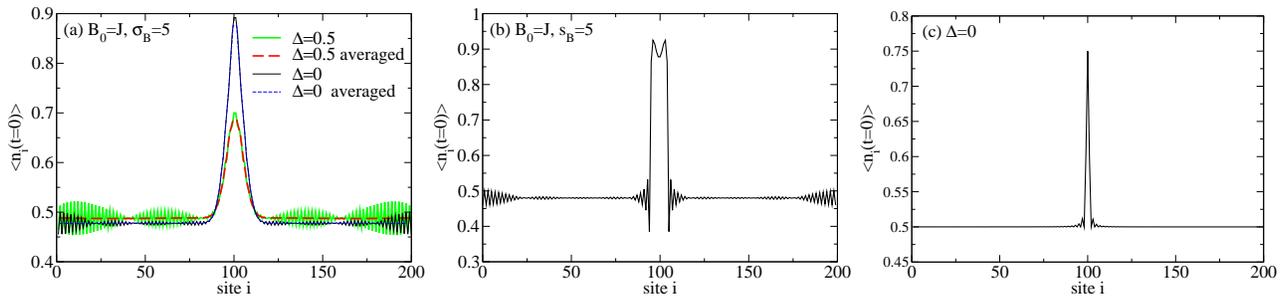

\includegraphics[width=0.31\textwidth,angle=0]{figure1a.eps}
\includegraphics[width=0.31\textwidth,angle=0]{figure1b.eps}
\includegraphics[width=0.31\textwidth,angle=0]{figure1c.eps}
\caption{(Color online) Density profiles $\langle n_i(t=0)\rangle=M_i(t)+1/2$ in the initial state
(time $t=0$),  induced by  (a) a Gaussian magnetic field Eq.~\eqref{eq:gauss}
($B_0=J, \sigma_B=5$); 
(b) a box-shaped magnetic field, Eq.~\eqref{eq:box} ($B_0=J$, $s_B=5$); and (c) application of
$S^+_{L/2+1}$, all at half filling $n=0.5$.
Thin solid lines in (a)-(c): $\Delta=0$; thick solid lines in (a): $\Delta=0.5$, thin and thick dashed lines in (a): averaged density $\langle \tilde n_i(t=0)\rangle$
for $\Delta=0$ and 0.5, respectively (see the text in Sec.~\ref{sec:init} for details on the averaging and Eq.~\eqref{eq:av}).
Results at $\Delta=0$ are obtained with ED, all others with DMRG.
}\label{fig:init}
\end{figure*}

\subsection{Methods}
For the time-evolution of the noninteracting case ($\Delta=0$ in Eq.~\eqref{eq:xxz_c}), we use exact diagonalization which allows
us to treat large systems. For all interacting cases, we employ the adaptive time-dependent DMRG method\cite{daley04,white04,vidal04}
with a Krylov-space-based time-evolution scheme\cite{park86,hochbruck97} in the space of matrix product states.\cite{mcculloch07} The control parameters are the discarded weight $\delta \rho$ and the time-step $\delta t$.
Our simulations are canonical ones as we work in subspaces with a fixed total $S^z$ or particle
 number, respectively.
We have performed an extended error analysis by (i) comparing ED and tDMRG results in the noninteracting case and (ii) by performing several runs 
with different time-steps and  discarded weights at representative 
parameters in the interacting case. Specifically, in case (i), we have analyzed the relative errors in the local magnetization 
$$
\delta M=\frac{1}{L}\sum_{i=1}^L|M_i^{DMRG}-M_i^{ED}|
$$ 
and its variance (see below). It turns out that typically, a time step of $\delta t=0.125/J$ and a discarded weight of $\delta \rho=10^{-6}$ keeps the relative error $\delta M$
in $M$ below $10^{-4}$ for a chain of $L=200$ sites at half filling and for times $t\leq 100/J$.

\section{Ballistic vs diffusive transport}
\label{sec:theory}
Within linear response theory, one often separates the dynamical conductivity $\sigma(\omega)$ into a 
delta-function $\delta(\omega)$ at zero frequency and a regular part at frequencies $\omega>0$,\cite{mahan,zotos97}
\begin{equation}
\sigma(\omega) = 2\pi D \delta(\omega) +\sigma_{\mathrm{regular}}(\omega) \,. \label{eq:drude}
\end{equation}
These quantities  derive from the Kubo formula that is based on evaluating current-current correlation
functions.\cite{mahan}
We repeat that a finite Drude weight in a clean, one dimensional system at zero temperature defines an ideal conductor and thus ballistic behavior, while if $D=0$, 
one has an insulator.\cite{kohn64,scalapino93}  
The dependence of the Drude weight on the exchange anisotropy $\Delta$ in the case of the integrable $XXZ$ chain and at zero temperature is well-known:\cite{shastry90}
\begin{equation}
D=\frac{\pi}{4} \frac{\sin \nu}{\nu(\pi-\nu)}  \,,\label{eq:drude_xxz}
\end{equation}
where, in this equation, the anisotropy is parameterized through $\Delta=\cos(\nu)$.
We thus have $D>0$ for $|\Delta| \leq 1$, featuring a discontinuous drop to zero at $\Delta=1$.
Due to the excitation gap in the massive regime $\Delta>1$, we have a true insulator with $\sigma_{\mathrm{dc}}=0$ that can only 
transport magnetization once the gap has been exceeded by a sufficiently large external perturbation.

Similarly, the Drude weight vanishes in the massive phases of both the spin ladder and the frustrated chain, while in the massless regime
of the latter model $D$ is  finite at $T=0$.\cite{bonca94,furukawa08} 

Strictly speaking, on all systems with open boundary conditions, the Drude weight vanishes identically. Yet, it turns out that the 
corresponding weight is just shifted to small but finite frequencies,\cite{kuehner00,rigol08} and thus the system is expected to 
still exhibit ballistic and anomalous transport properties.

To justify and motivate our way of analyzing ballistic and diffusive transport, let us,
for pedagogical reasons, consider a 1D system that
obeys the diffusion equation:
\begin{equation}
\partial_t \rho(x,t)=\nabla\cdot (\mathcal{D}\nabla \rho(x,t))\,.
\end{equation}
Here, $\rho(x,t)$ denotes, {\it e.g.},  a particle density, and $\mathcal{D}$  the diffusion constant.
The Green's function associated with this equation is, in a $d$-dimensional setup, given by
\begin{equation}
G(x,\acute{x},t)=\frac{1}{(4\pi \mathcal{D}t)^{d/2}}\cdot e^{-\frac{(x-\acute{x})^2}{4\mathcal{D}t}}\,.
\end{equation}
Therefore,  we can calculate the expectation values
\begin{equation}
\langle x \rangle (t)= \acute{x} \ \mbox{and} \ \langle x^2\rangle=|\acute{x}|^2+2\,d \mathcal{D}t
\end{equation}
and see that the variance $\sigma_x^2 =\langle x^2\rangle- \acute{x}^2$ is linear in $t$ for normal diffusive transport. On the contrary, for ballistic
dynamics, one expects the variance $\sigma_x^2$ to grow quadratically in time, as is well-known from elementary quantum mechanics for
free particles.

Given a distribution of $M_i(t)=\langle n_i(t)\rangle -1/2$ at a time $t$, we find it most straightforward to compute the variance
from the corresponding particle density distribution $\langle n_i(t) \rangle=M_i(t)+1/2$ as this is a positive quantity.
We then compute the variance from
\begin{equation}
\sigma_M^2(t)= \frac{1}{(L/2)}\sum_{i=1}^{L}(i-\mu_n)^2\cdot \langle n_i(t)\rangle \,.\label{eq:sigma}
\end{equation}
where $\mu_n$ is the first moment of the normalized distribution $\langle n_i\rangle$. Note that we normalize
$\langle n_i\rangle$ on the actual number of fermions rather than the system size.

\begin{figure}[b]
\includegraphics[width=0.45\textwidth,angle=0]{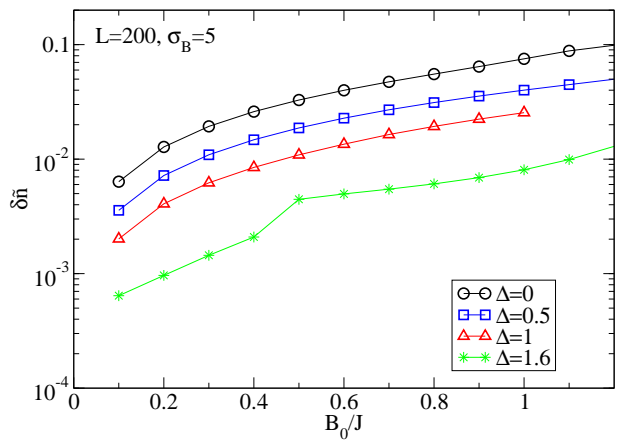}
\caption{(Color online) Average deviation $\delta \tilde n$ from half filling [see Eq.~\eqref{eq:deltan}] in the initial state prepared by applying  
a Gaussian magnetic field [Eq.~\eqref{eq:gauss}] with $\sigma_B=5$ as a function of $B_0$ for
$\Delta=0,0.5,1$ and $1.6$. 
}\label{fig:delta_n}
\end{figure}

\begin{figure}[t]
\centerline{\includegraphics[width=0.45\textwidth,angle=0]{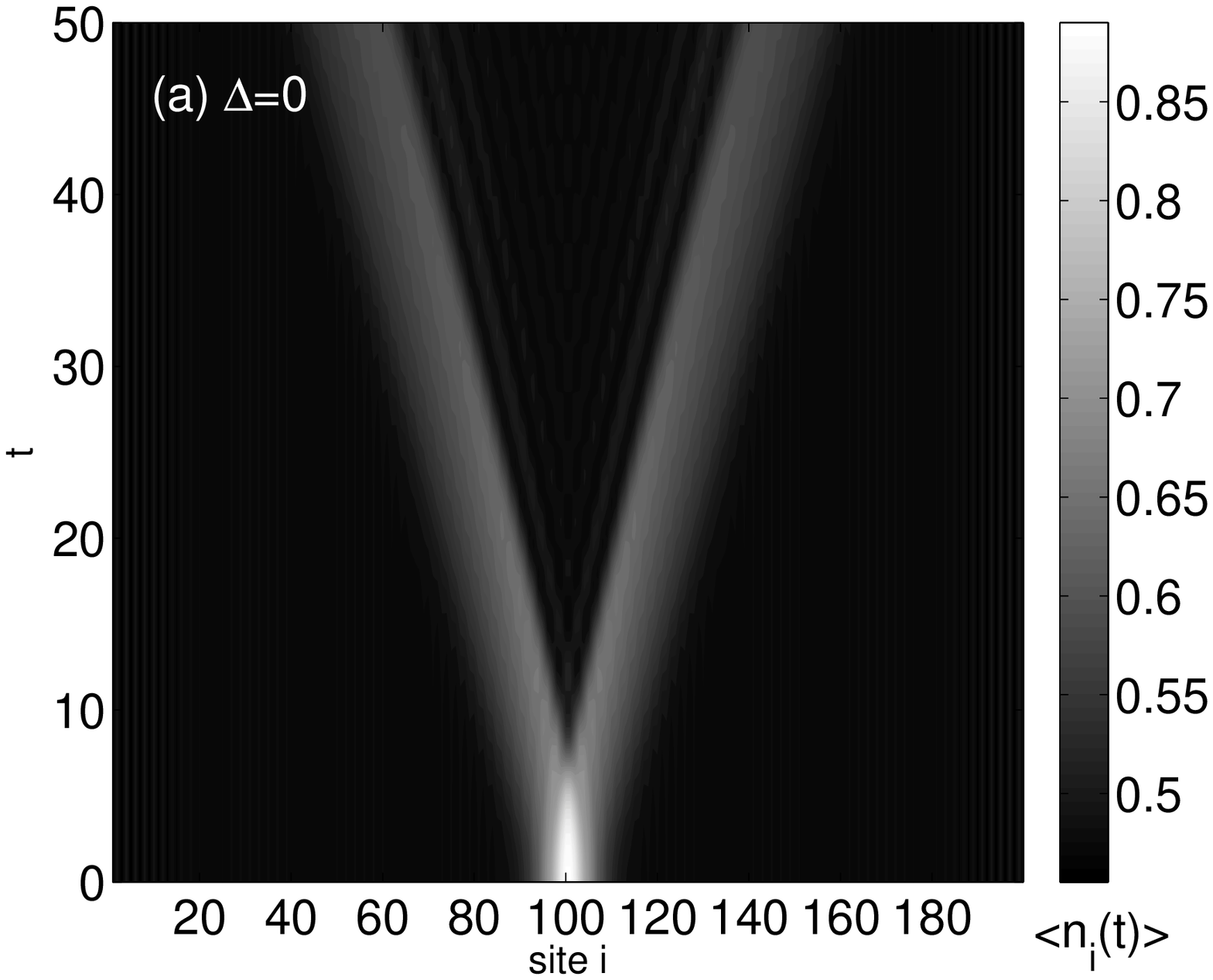}}
\centerline{\includegraphics[width=0.45\textwidth,angle=0]{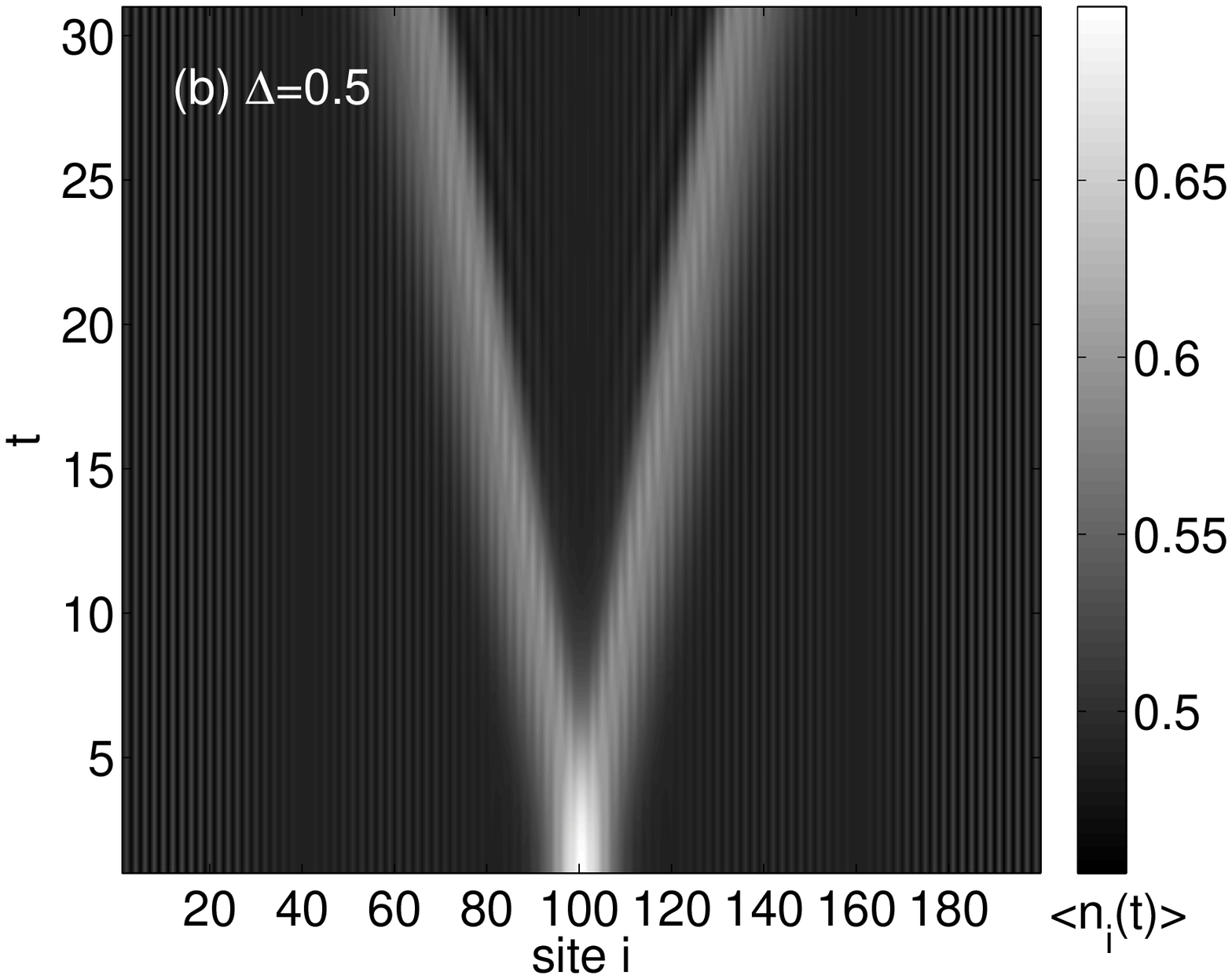}}
\caption{(a), (b)
Time-evolution of  the density (magnetization) profile $\langle n_i(t)\rangle=M_i(t)+1/2$
for a Gaussian initial state with  $B_0=J$, $\sigma_B=5$  for (a): $\Delta=0$ (ED data) and (b): $\Delta=0.5$ (DMRG data, $L=200$ sites).
}
\label{fig:U0_states}
\end{figure}


\section{Preparation of initial states}
\label{sec:init}

We consider three different ways of preparing the initial state, which we now illustrate 
for the case of the $XXZ$ chain, Eq.~\eqref{eq:xxz}, and at half filling $n=0.5$, {\it i.e.}, at $S^z=0$. 
With the exception of Sec.~\ref{sec:restore}, our simulations are always performed in subspaces with these
quantum numbers.

Technically, we add a term 
\begin{equation} 
H_{B} = -\sum_i B_i S_i^z \label{eq:h_b} 
\end{equation}
to the Hamiltonian. By choosing the $B_i$ appropriately,  this realizes (i) a Gaussian magnetic field, (ii) a box-shape magnetic field,
both applied during a ground-state DMRG run, and turned off at time $t=0$:
\begin{eqnarray}
B_i(t=0) &=& B_0\, \mbox{exp}(-(i-i_0)^2/2\sigma_B^2)\label{eq:gauss}\\
B_i(t=0) &=& B_0 \, \Theta\left(i-\frac{L-s_B}{2}\right) \Theta\left(i+\frac{L+s_B}{2}\right)   \label{eq:box} \,.
\end{eqnarray}
$\Theta(i)$ is the Heaviside function.
In Eq.~\eqref{eq:gauss}, $\sigma_B$ is the variance of the external Gaussian field, $B_0$ its amplitude, and we set $i_0=L/2+0.5$,
while in Eq.~\eqref{eq:box}, $s_B$ denotes the width of a box in which a constant field $B_0$ is applied.

 The third initial state (iii) is realized by finding the
ground state of a system at filling $n=(L/2-1)/L$ and then applying a single spin flip $S^+_i$ on a site $i$. The time-evolution is then 
performed at half filling, as in cases (i) and (ii).

The typical shape of the induced density $\langle n_i(t=0)\rangle $, or equivalently, the  magnetization profile 
 is illustrated in Figs.~\ref{fig:init}(a)--(c) at $\Delta =0$
(all panels) and $\Delta=0.5$ (panel (a) only], for the three initial states (i)--(iii), respectively.
Inherent to the fermionic nature of the model Eq.~\eqref{eq:xxz_c}, we first observe Friedel oscillations with a $2k_F$ period, where $k_F=\pi/2$ 
is the Fermi momentum at half filling.
Second, there are slower spatial oscillations, that are more evident in the case of $\Delta=0.5$ [Fig.~\ref{fig:init}(b)].
These oscillations' characteristic wave-length depends, as we have checked, on $\sigma_B$ as well as the system size.
As we shall see later, for the purpose of qualitatively analyzing the time-dependence of the variance, the presence of the
long-ranged oscillations is irrelevant, as the dynamics stems from the cental peak dispersing, while away from the center, the
oscillations contribute subdominantly to the time-dependence after turning off $H_{B}$. We will nevertheless sometimes find it 
illustrating and useful to work with a density \begin{equation}
\langle \tilde n_x(t)\rangle =  \lbrack\langle n_{2i-1}(t)\rangle + \langle n_{2i}(t)\rangle\rbrack/2
\,\label{eq:av}
\end{equation}
averaged over adjacent sites with $i=1,\dots,L/2$ and $x=2i-1/2$. To recover the variance of the non-averaged density, we multiply $\tilde \sigma_M^2$ by a factor of $2$.
 The averaged density $\langle \tilde n_i(t=0)\rangle$ is plotted with dashed lines in
Fig.~\ref{fig:init}(a) for $\Delta=0$ and 0.5, and 
we see that this averaging results in quite smooth curves.

For $\langle \tilde n_i(t=0)\rangle$ and a Gaussian external $B_i$, we can now further address the question whether the induced density profile follows a Gaussian as well.
We find that this is  the case at small $\Delta$, in good approximation. In the massive phase $\Delta>1$, deviations of $\langle \tilde n_i(t=0)\rangle$ from
a Gaussian profile are substantial. We will nevertheless refer to initial states prepared with
Eq.\eqref{eq:gauss} as Gaussian initial states throughout.

As we go from $\Delta =0$ into the massive regime $\Delta>1$, a gap opens, and qualitatively, despite the field $B_i$ being inhomogeneous,
we expect the existence of the gap to affect the average deviation from half filling at a given set of $(B_0,\sigma_B)$.
To illustrate this point, we display this average deviation $\delta \tilde n$, defined as
\begin{equation}
\delta\tilde n = \sqrt{\sum_i (\langle \tilde n_i\rangle -1/2)^2/N}\,,\label{eq:deltan}
\end{equation}
in Fig.~\eqref{fig:delta_n} ($\Delta=0,0.5,1,1.6$). For $\Delta=1.6$, we observe a steep increase in $\delta \tilde n$ at $B_0/J\approx 0.4$, and we will
use $B_0/J>0.45 $ for the time-evolution, as below this value, little dynamics in the time-evolution is seen. Note that Fig.~\ref{fig:delta_n}
has a logarithmic scale on the $\delta \tilde n$ axis, and below $B_0/J\approx 0.4$, $\delta \tilde n \propto {\mbox{exp}}(c/B_0)$ in the case of $\Delta=1.6$.
We find $c=0.4J$ and this roughly coincides with twice the spin gap for this value of
$\Delta$.\cite{cloizeaux66}

\begin{figure}[t]
\centerline{\includegraphics[width=0.45\textwidth,angle=0]{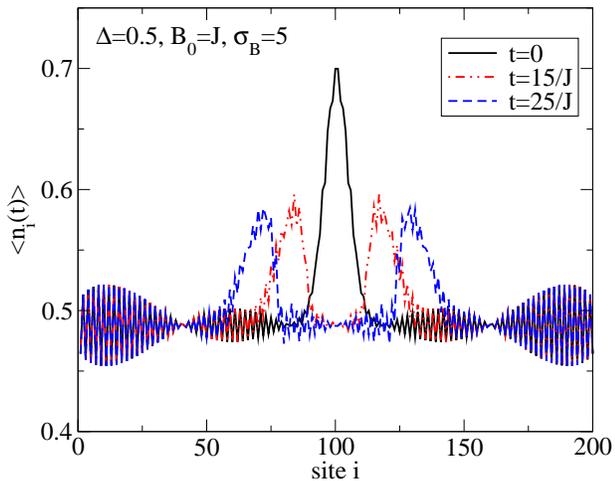}}
\caption{(Color online) 
Time-evolution of  the density $\langle n_i(t)\rangle$ (or magnetization profile $M_i(t)=\langle n_i(t)\rangle-1/2$)
for a Gaussian initial state with  $B_0=J$, $\sigma_B=5$ at $\Delta=0.5$ (DMRG data, $L=200$ sites):
  Snap-shots from Fig.~\ref{fig:U0_states} at times $t\,J=0,15,25$.}
\label{fig:snap}
\end{figure}


\begin{figure*}[t]
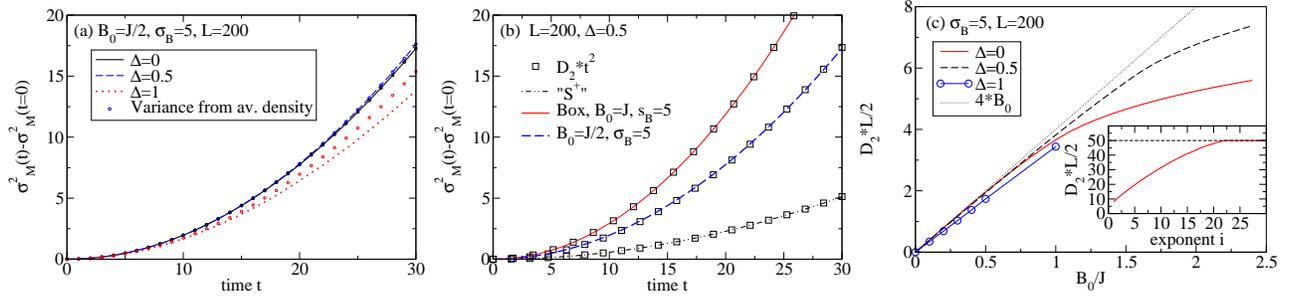

\includegraphics[width=0.31\textwidth,angle=0]{figure5a.eps} 
\includegraphics[width=0.31\textwidth,angle=0]{figure5b.eps}
\includegraphics[width=0.31\textwidth,angle=0]{figure5c.eps}
\caption{(Color online) 
(a) Time-dependence of the variance $\sigma_M^2(t)-\sigma_M^2(t=0)$ for $\Delta=0$ (solid lines) and $\Delta=0.5$ (dashed lines), both at $B_0=J/2$ 
and $\sigma_B=5$. Circles denote the variance  $\tilde\sigma_M^2(t)-\tilde\sigma_M^2(t=0)$ of
the averaged density $\langle \tilde n_i(t)\rangle $, Eq.~\eqref{eq:av}. 
(b) Time-dependence of the variance $\sigma_M^2(t)-\sigma_M^2(t=0)$ for the evolution from a box-like initial state (solid line), 
after the application of $S^+_{L/2}$  (dot-dashed line), and from a Gaussian-state (dashed line) (all at $\Delta=0.5$). The squares denote the 
perfect fit of all sets to $\sigma_M^2(t)-\sigma_M^2(t=0) = D_2 t^2$. 
(c) $B_0$-dependence of the  coefficient $D_2$  of the time-dependent variance  for $\Delta=0$ (solid lines), $\Delta=0.5$ (dashed lines), and $\Delta=1$ (circles)
at a fixed $\sigma_B=5$ and $L=200$.  The dotted lines display $D_2\,L/2=4 B_0J$. Inset: $D_2(B_0)$ in the 
limit of large $B_0=10^{i}J$ for $\Delta=0$. The thin dashed line is the result for the  expansion from a
Fock state of width $L/2$ with $\langle n_i\rangle=1 $  at the center of the chain. }
\label{fig:U0_DofB}
\end{figure*}

\section{The $XXZ$ chain}
\label{sec:xxz}
After detailing the way of preparing initial states, we now come to the analysis
of $M_i(t)=\langle n_i(t)\rangle-1/2$, focusing on the variance $\sigma^2_M(t)$. 
In this section, we will first discuss $\sigma_M^2(t)$ in the massless regime in Sec.~\ref{sec:massless}, and show that we clearly
see ballistic behavior with  $\sigma_M^2(t)=\mbox{const}+D_2 t^2$, independently of the initial state.
We will then analyze the dependence of the coefficient $D_2$ on $B_0$, 
scanning the full range of perturbations from a linear one with $D_2\propto B_0$ to the
 largest perturbations possible. These regimes are distinct by a different finite-size scaling behavior, to be discussed below.
As we illustrate in the case of a Gaussian magnetic field Eq.~\eqref{eq:gauss}, 
 ballistic dynamics is found for $\Delta\leq 1$. This holds independently of the actual choice of $(B_0,\sigma_B)$
and thus also far from equilibrium, except for the most extreme initial states considered here (see the discussion in Sec.~\ref{sec:massless_b} below).

We then, in Sec.~\ref{sec:xxz_massive}, discuss the transition from ballistic to diffusive behavior which is expected to occur at $\Delta=1$.
Our data are consistent with this picture, as we find strong evidence for diffusive transport for $\Delta \gtrsim 1.5$. We will present results
for several $B_0$ at $\Delta=1.5$ to substantiate that the observation of diffusive transport is independent of the initial state.

\begin{figure}[b]
\includegraphics[width=0.31\textwidth,angle=0]{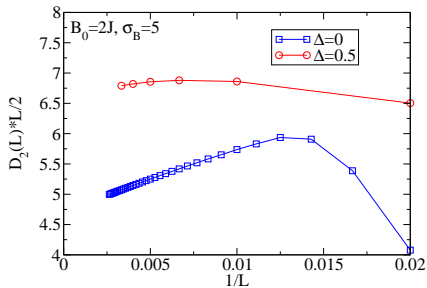} 
\caption{(Color online) 
Finite-size scaling of $D_2$ for $\Delta=0$ and 0.5 in the large $B_0$ regime ($B_0=2J$, $\sigma_B=5$).}
\label{fig:fs}
\end{figure}

\subsection{$XXZ$ chain: Massless regime} \label{sec:massless} 

\subsubsection{Time-dependence of
the variance} We now turn to the analysis of $M_i(t)=\langle n_i(t)\rangle -1/2$ in the massless
regime. Figures~\ref{fig:U0_states}(a) and (b) show $\langle n_i(t)\rangle=M_i(t)+1/2$ for
$\Delta=0$ and $\Delta =0.5$, 
respectively. The initial density profile first melts and then,  at times $t\,J\approx 5 $, splits
into two packets that travel into opposite directions.\cite{kollath05b}   Figure~\ref{fig:snap}
shows snapshots of $\langle n_i(t)\rangle$ at times $t\,J=0,15,25$ for $\Delta=0.5$. It is
noteworthy that, while substantial oscillations are present far away from the central  peak, these
oscillations are frozen in and do not contribute to the increase in $\sigma_M^2$ since, far away from the
center of the chain,
$H\approx H+H_B$. 

The variance $\sigma_M^2(t)-\sigma^2_M(t=0)$, plotted vs.~time, is displayed in Fig.~\ref{fig:U0_DofB}(a), for $\Delta=0$ (solid line),  $\Delta=0.5$ (dashed line), and $\Delta=1$ (dotted line)
for the evolution from an initial state of the type (i), enforced by a Gaussian magnetic field with
$B_0=J/2$ and $\sigma_B=5$.
The circles represent the time-evolution of $\tilde \sigma^2_M(t)-\tilde \sigma^2_M(t=0)$ 
computed from the averaged density $\langle \tilde n(t)\rangle $ 
(see Sec.~\ref{sec:init}) for $\Delta=0,0.5,1$. 
For $\Delta=0$ and $0.5$, the averaged results very well coincide with $\sigma^2_M(t)$. 
For the purpose of characterizing ballistic or
diffusive behavior, it therefore does not matter whether the pure fermionic density $\langle n_i(t)\rangle$ or the averaged quantity $\langle \tilde n_i(t)\rangle$
is used. In what follows, we will present results extracted from the former, unless stated otherwise. We mention, though, that the quantitative difference in the variance  extracted
from the averaged as compared to the bare density becomes 
more pronounced the larger $\Delta$ and the smaller $B_0$ is. This becomes evident in the case of
$\Delta=1$, included in Fig.~\ref{fig:U0_DofB}(a).

The key observation from Figs.~\ref{fig:U0_DofB}(a) and (b) is the quadratic increase of the variance with time observed for $\Delta=0,0.5$ and 1, which 
confirms the expected ballistic behavior in the critical regime. 
For the isotropic chain ($\Delta=1$), we find that the best fit of  a power-law  to $\sigma^2_M(t)-\sigma^2_M(t=0)$ yields $  \sigma^2_M(t)-\sigma^2_M(t=0) \propto t^{1.98}$,
 which is thus slightly below the behavior expected for ballistic transport. However, a  deviation of just one percent from the expected exponent of 2 is very much within the accuracy of our numerical calculations. 

Moreover, this behavior, as we show in Fig.~\ref{fig:U0_DofB}(b) for the example of $\Delta=0.5$, 
is independent of
the shape of the initial state: all three types of states studied here - Gaussian field, box shape, and application of $S_i^+$ - result
in ballistic dynamics at half filling. 

\begin{figure*}[t]
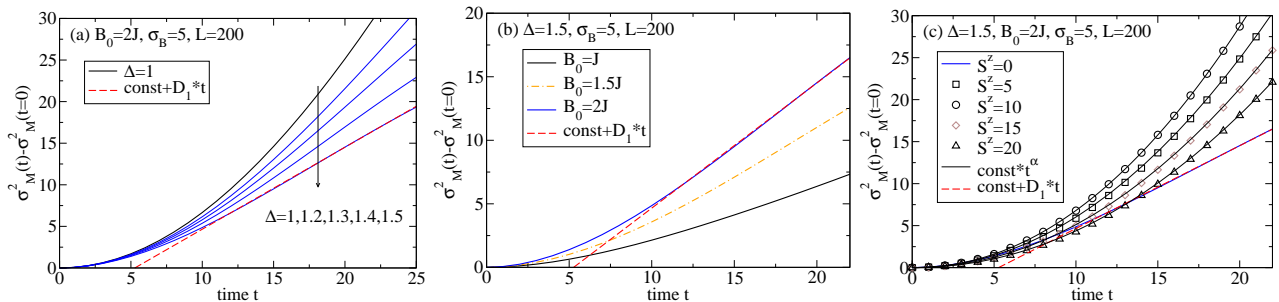

\includegraphics[width=0.31\textwidth]{figure7a.eps}
\includegraphics[width=0.31\textwidth]{figure7b.eps}
\includegraphics[width=0.31\textwidth]{figure7c.eps}
\caption{(Color online) (a) Transition from ballistic to diffusive transport: 
$\sigma^2_M(t)-\sigma^2_M(t=0)$ for   $\Delta=1$ (thick solid line) $\Delta= 1.2,1.3,1.4,1.5$ (thin  solid lines). 
The parameters are $B_0=2J$, $\sigma_B=5$, $L=200$. The thick dashed line shows the asymptotic behavior for
$\Delta=1.5$, {\it i.e.}, $\sigma^2_M(t)=\mbox{const}+D_1t$. 
(b) Variance   for $\Delta=1.5$ and several $B_0/J=1,1.5,2$ ($L=200$, $\sigma_B=5$).
(c) Variance for $\Delta=1.5$  at $S^z=0,5,10,15,20$ (see the legend) and $B_0=2J$, $\sigma_B=5$: ballistic transport is restored away from half-filling.
}
\label{fig:delta}
\end{figure*}

\subsubsection{From small to large perturbations}
\label{sec:massless_b}
So far, we have worked at a fixed pair of $B_0$ and $\sigma_B$. We will now explore the dependence of the dynamics on how far 
the initial state is perturbed away from the actual ground state. 
This deviation is measured through  the average deviation $\delta \tilde n$
from half filling (compare Fig.~\ref{fig:delta_n}).
For that purpose, we fix $\sigma_B=5$, and perform a set of simulations at different $B_0$, at
$\Delta=0,0.5$ and 1.
Independently of $B_0$, we always find a quadratic increase of $\sigma_M^2(t)$ with time, {\it
i.e.}, $\sigma_M^2(t)=\mbox{const}+D_2(B_0) t^2$. As a main result, we therefore
conclude that the {\it qualitative} behavior of the dynamics is independent of the external perturbation, it is always ballistic in the massless regime.

On a more quantitative level, it is instructive to plot $D_2$ vs $B_0$, 
shown in Fig.~\ref{fig:U0_DofB}(c) for $\Delta=0$ and 0.5, as it allows us to distinguish two
regimes: first, the linear regime, in which $D_2$ is linear in $B_0$ with $D_2\,L/2\not=f(L)$ at a fixed $B_0$.
For $\Delta=0$ and 0.5, the linear regime extends up to $B_0/J \approx 0.5$. 
Second, at larger $B_0$ effects of both the band curvature and the finite band-width 
start to play a role, with significant finite-size effects, as illustrated in Fig.~\ref{fig:fs} 
for $B_0=2J$.
In the case of $\Delta=0$ and $B_0=2J$, we are able to access system sizes of $L\approx 10^{5}$, and at large
$L$, the scaling is of the form $D_2\, L/2\propto 1/L$, allowing for an extrapolation to $L\to\infty$. In the
interacting case, the accessible system sizes are too small to establish such scaling and we thus have not
attempted any extrapolation in the case of $\Delta=0.5$.

Let us next discuss the limiting cases of first, the linear regime, {\it i.e.}, $D_2\propto B_0$, and second, the limit of $B_0\to \infty$.
Starting with the former, the linear regime, $D_2 \,L/2= \gamma B_0$, we find that the prefactor is $\gamma\approx 4J$ for  values of $\Delta \lesssim 0.5$.
At large $\Delta$, this reduces to $\gamma\approx 3.4 J$ as the results for $\Delta=1$ displayed in
Fig.~\ref{fig:U0_DofB}(c) show (circles).
The interpretation of $D_2$ being linear in $B_0$ for $\Delta\leq 1$ is based on the observation that the 
area $A_{\mathrm{peak}}(B_0)$ under the initial Gaussian-like magnetization profile increases 
linearly in $B_0$, which we can strictly confirm in the $\Delta=0$ case and chains with up to $L=1000$ sites. This analysis requires an estimate
of the background density, which can the best be done in the $\Delta=0$ case, but suffers from finite-size effects at a nonzero $\Delta$.
In the $\Delta=0$ case, we find that $\gamma B_0/A_{\mathrm{peak}}(B_0) =4 J^2$ and for $\Delta=0.5$, $\gamma B_0/A_{\mathrm{peak}}(B_0) \approx 1.7 J^2$.
We may therefore conclude that in the linear regime,  $\gamma\,B_0=A_{\mathrm{peak}}(B_0) \,v_{g}^2(\Delta)$, where
$v_{g}(\Delta)$ is the group velocity (see, {\it e.g.}, Ref.~\onlinecite{cabra-review}):
\begin{equation}
v_{g}(\Delta)= \frac{\pi}{2} \frac{\sin(\nu)}{\nu}
\end{equation}
with $\Delta=\cos(\nu)$.
From our numerical data for $L=200$ sites and at $\Delta>0.5$, we obtain $D_2\, L/(2A_{\mathrm{peak}})>v_{g}^2(\Delta)$ due to finite-size effects, but 
qualitatively, $D_2 \,L/(2A_{\mathrm{peak}})$ increases with $\Delta$ at a fixed $B_0$ in the linear regime as expected from Eq (15) for the behavior of the velocity $v_g^2$. 
At sufficiently large $B_0$, we expect the band-curvature to play a role as well which we are able to verify in the case of $\Delta=0$. $D_2
\,L/(2A_{\mathrm{peak}})$
then decreases below $v_g^2(\Delta=0)$ as $B_0$ increases. 

As for the limit of large $B_0\gg J$, note that since $M_i\leq 1/2$ and since we work at fixed  half filling of the full system, the most extreme
initial state that, on a fixed system size $L$, $B_0\to \infty$ drives the system into  is  a Fock state $| f \rangle $ with 
$\langle n_i(t=0)\rangle=1$ for $L/4< i< 3L/4$ and zero otherwise. We  thus follow the evolution from such a state as well as the evolution from 
states with very large $B_0\sim 10^{i}J$ ($i=1,\dots,30$). The inset in Fig.~\ref{fig:U0_DofB}(c) 
shows that as $B_0$ increases, $D_2(B_0)$ indeed approaches 
the value found for  the limiting state $| f \rangle $ for $B_0>10^{22}J$ in the example of $\Delta=0$. 
In that limit, $D_2=1/2 J^2$. 

While for the parameters of the main panel of Fig.~\ref{fig:U0_DofB}(c), {\it i.e.}, $\Delta=0,0.5,1$ and $B_0/J\leq 2.4$, 
the variance always follows a power law with exponent two,   curiously, this is not the case for the aforementioned Fock states encountered in the
$B_0\to \infty$ limit and $\Delta=0.5,1$. There, we find  exponents that are consistently below two for $\Delta>0$ and $L=40,80,200$.
This behavior was also observed in Ref.~\onlinecite{gobert05} for the evolution from similar Fock states.
A full analysis of the evolution from Fock states will be presented elsewhere.

We mention that the time-dependent evolution from Fock states or, more generally, the evolution of particles originally trapped in 
a confined region into an {\it empty} lattice, 
has been intensely studied with numerical methods in Refs.~\onlinecite{rigol04,rigol05b,rodriguez06,hm08a},
for the cases of hardcore bosons \cite{rigol04,rigol05b},
soft-core bosons,\cite{rodriguez06} and the interacting two-component Fermi gas.\cite{hm08a} These studies have a fundamental interest in out-of-equilibrium phenomena,
with a  perspective onto experiments with ultra-cold atoms in optical lattices. Among these we mention the experimental efforts directed at detecting
Anderson localization in cold atom gases, precisely by utilizing such expansion set-ups.\cite{clement05,fort05}

Finally, while all results discussed here were obtained from chains with $L=200$, 
we stress that we have carefully
studied the finite-size scaling of $D_2$(see Figs.~\ref{fig:U0_DofB}(c) and\ref{fig:fs}). By plotting $D_2\,L/2$ in Fig.~\ref{fig:U0_DofB}(c), we account for  a trivial size-dependence,
and curves obtained from  different $L$ but the same $\Delta$ indeed collapse onto a single one  in the linear regime. At larger $B_0$,
$D_2$ tends to decrease with system size $L$ as less particles per $B_0$ can be 
accumulated in the initial Gaussian peak, due to $S_i^z\leq 1/2$ (see Fig.~\ref{fig:fs}).


\subsection{$XXZ$ chain: Massive phase}
\label{sec:xxz_massive}
\subsubsection{Transition from ballistic to diffusive behavior}
Linear-response theory predicts a sharp transition from ballistic spin transport, characterized by a finite Drude weight $D$,
to diffusive spin transport at $\Delta=1$.\cite{shastry90}
We have carried out several simulations with different $\Delta=1,1.2,1.3,1.4,1.5$ at fixed parameters $B_0=2J$ and $\sigma_B=5$
to check whether an analysis of the variance captures this transition. As the data for $\sigma_M^2(t)$
displayed in Fig.~\ref{fig:delta}(a) show,
we clearly find a linear increase of  $\sigma_M^2(t)$ at large times in the case of $\Delta=1.5$, which, in the sense of Sec.~\ref{sec:theory},
we interpret as evidence for diffusive transport. 
At smaller $1<\Delta\lesssim 1.4$, our data do not allow for this conclusion, yet at least we can state that the data for $\Delta=1.3$ and $1.4$ do not follow a power-law,
indicative of non-ballistic transport.
 Also note that there is no  contradiction with 
Eq.~(\ref{eq:drude_xxz}) as it is is well-known that finite-size effects in the vicinity of $\Delta=1$ are severe and come along with a logarithmically slow
convergence with system size of quantities such as the Drude weight or the spin stiffness.\cite{laflorencie01,hm03} We suspect that larger system sizes, hence access to 
longer simulation times, are necessary to fully capture the sharp transition from ballistic to diffusive
behavior at $\Delta=1$.
We stress that with $B_0=2J$, we work with highly perturbed initial states, and thus 
the observation of diffusive transport for $\Delta\geq 1.5$ is a nontrivial one, going beyond
the case studied in Ref.~\onlinecite{zotos97}.
Our results for $\Delta\geq 1.5$ thus establish an example of diffusive dynamics with $\sigma_M^2(t)=\mbox{const}+D_1 \,t$  in this model for the out-of-equilibrium situation.
Note that a recent tDMRG work on transport in spin chains incorporating baths has reported similar results, derived from current and spin profiles in
the steady state.\cite{prosen09}

It is noteworthy  that at short times, obviously, the dynamics is always ballistic, independently of $\Delta$, as can be seen in  Fig.~\ref{fig:delta}(a).
Even the value of $\sigma_M^2(t)$ is roughly the same for all $\Delta$ at short times. 
In the long time limit, which   is the relevant one to characterize the 
system as diffusive or ballistic, we find that $\sigma_M^2(t)$ systematically decreases with increasing $\Delta$. The reason is that, in the $\Delta/J\to \infty$ limit,
no dynamics is possible at all.

We have further studied the dependence on $B_0$ at $\Delta=1.5$.
In Sec.~\ref{sec:init} we hinted at the fact that at $\Delta>1$, a reasonably large $B_0$ is necessary to observe a significant change in the 
magnetization profile over the time scales simulated. 
We attribute this to the existence of a spin gap in the antiferromagnetic phase $\Delta>1$
and focus on $B_0\gtrsim J$.
Our results of several runs  at $\Delta=1.5$, scanning the $B_0$ dependence, are displayed in Fig.~\ref{fig:delta}(b). We find that the variance increases with 
$B_0$, which we mainly attribute to more particles accumulated in the central peak. Furthermore, the figure
suggests that the time-scale at which diffusive behavior sets
in depends on $B_0$ such that for larger $B_0$,  we are able to observe a linear increase of  $\sigma_M^2(t)$ earlier in time.

\subsubsection{Restoring ballistic transport}
\label{sec:restore}
While so far we have restricted ourselves to the case of half filling, we now
address the magnetization dynamics at incommensurate filling. 
The initial states are now created by applying the external fields in subspaces 
that already have a finite magnetization., {\it i.e.} $S^z\not= 0$.
Results for the variance 
at $\Delta=1.5$ are displayed in Fig.~\ref{fig:delta}(c). The $z$-component of the total spin is $S^z=0,5,10,15,20$.
We find that any nonzero $S^z$ is sufficient to render the dynamics
ballistic again, on the time scales accessible to our simulations. 
According to our data, the variance follows $\sigma_M^2(t)-\sigma_M^2(t=0)\propto t^{\alpha}$, with
$\alpha\approx 2.05\pm 0.02$.
This observation is in agreement
with the infinite-temperature behavior of the Drude weight,\cite{zotos97,hm05} which in the $XXZ$ chain
is finite at any $\Delta$ away from half filling. Note that 
$\Delta>1$ and $S^z>0$ but below saturation is in the easy-plane phase
of the $XXZ$ chain, with gapless excitations.\cite{review-frust} Note that the prefactor in $\sigma_M^2(t)-\sigma_M^2(t=0)=D_2(S^z) t^{\alpha}$
depends on $S^z$ in a non-monotonous way: it is the largest at around $S^z=10$ and then decreases as saturation is reached.

\section{Nonintegrable models}
\label{sec:non}

We finally move on to the discussion of the dynamics in two nonintegrable models,
the two-leg ladder and the frustrated chain, both limiting cases of 
Eq.~\eqref{eq:non}. Numerical studies of the {\it high}-temperature limit, based on the Kubo formula, conclude that spin and thermal transport
in the massive phases of these models (see Sec.~\ref{sec:model}) are normal, with a vanishing Drude weight.\cite{hm03,hm04,hm07a,zotos04}
The conclusions on the massless phase of the frustrated chain are not unambiguous,\cite{hm03,jung06,hm07a,jung07a} and it has been pointed out
that the {\it energy-current operator}, 
to first order in the next-nearest-neighbor interaction $\alpha J$, is conserved.\cite{jung06} 

\begin{figure}[t]
\includegraphics[width=0.45\textwidth]{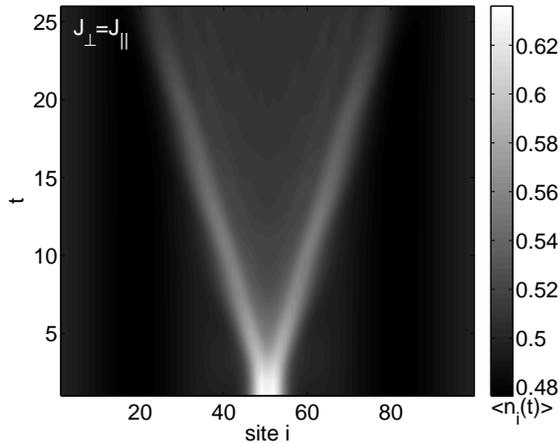}
\caption{Two-leg spin ladder with $J_{\parallel}=J_{\perp}$ [$\alpha=1$, $\delta=1$ in
Eq.~\eqref{eq:non}]:
 Contour plot of $\langle n_i(t)\rangle=M_i(t)+1/2$. 
In the initial state, a Gaussian external field $B_i$ with $B_0=J$ and $\sigma_B=5$ is applied [see Eq.~\eqref{eq:gauss}].
DMRG data, $L=200$ sites.
}
\label{fig:ladder}
\end{figure}

One scenario is that the high-temperature Drude weight vanishes,\cite{hm03,hm04} while it is still possible to find {\it anomalous}
transport properties in the low-temperature regime, {\it e.g.}, in the form of a peculiar low-frequency behavior of $\sigma_{\mathrm{regular}}(\omega)$.
Exact diagonalization results show that the Drude weight is finite at zero temperature in the massless phase
of the frustrated chain.\cite{bonca94,furukawa08}

We here use the approach outlined  in the previous sections to  show that the zero-temperature dynamics
of two-leg ladders and frustrated chains with a spin gap is of diffusive nature. To this end, we prepare initial states
with Gaussian magnetic fields Eq.~\eqref{eq:gauss}. We emphasize that, in the case of the spin ladder, both sites on a rung experience the same 
field. 
 In these two cases, and similar
to the discussion of initial states for $\Delta>1$ in the $XXZ$ chain (compare Fig.~\ref{fig:delta_n} in Sec.~\ref{sec:init}),
the amplitude of the Gaussian field, $B_0$, needs to be large enough to induce a substantial perturbation in the magnetization $M_i$ 
that will actually
propagate  through the system. We thus here probe the magnetization dynamics and transport at large external perturbations.

\begin{figure}[t]
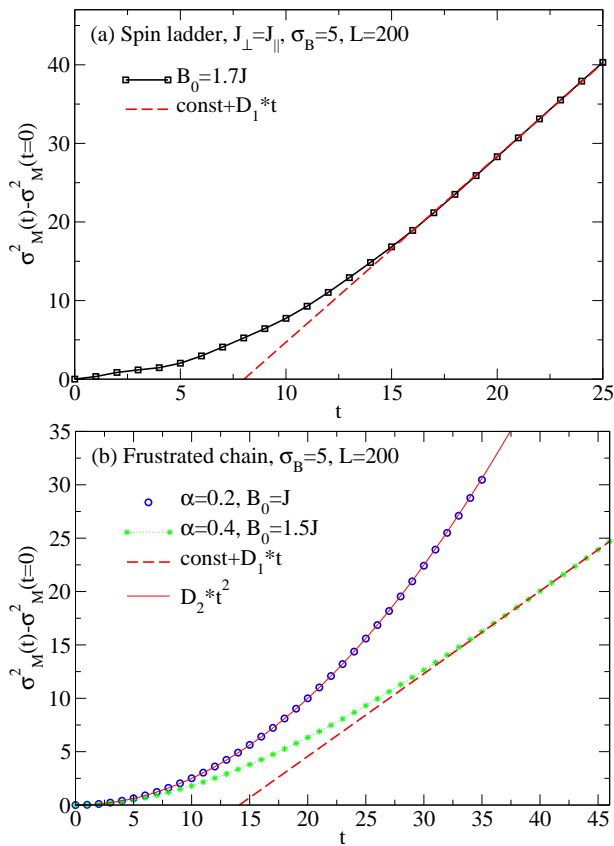

\includegraphics[width=0.45\textwidth]{figure9a.eps}
\includegraphics[width=0.45\textwidth]{figure9b.eps}
\caption{(Color online) Time dependence of the variance $\sigma^2_M(t)-\sigma^2_M(t=0)$. 
(a) Spin ladder, solid line with squares, compiled from the data of Fig.~\ref{fig:ladder}, $B_0=1.7J$. 
(b) Frustrated chain with $\alpha=0.2$ (circles, $B_0=J$) and a frustrated chain with $\alpha=0.4$ (stars, $B_0=1.5J$).
In all cases, $\sigma_B=5$.
The dashed lines in (a) and (b) are a fit of const$+D_1\, t$ to the variance of the spin-ladder and the frustrated chain with $\alpha=0.4$ at long times, 
while the thin, solid line in (b) is a least-square fit of $D_2\, t^2$ to the result for the frustrated  chain with $\alpha=0.2$.
DMRG data, $L=200$ sites.
}
\label{fig:ladder_v}
\end{figure}

Starting with the example of a spin ladder with $J_{\perp}=J_{\parallel}$, 
we display the magnetization profile $M_i(t)=\langle n_i(t)\rangle-1/2$ in Fig.~\ref{fig:ladder} as a contour plot.
The time-dependence of the corresponding variance is shown in Fig.~\ref{fig:ladder_v}(a) [solid line,
squares], and we find a linear 
increase in $\sigma_{M}^2(t)$ for times $t\gtrsim 17/J$, clearly establishing the notion of diffusive dynamics in the ladder system.

A more involved  picture emerges in the case of  the frustrated chain.
For this model we present results for $\alpha=0.2$ (circles) and $\alpha=0.4$ (stars) in Fig.~\ref{fig:ladder_v}(b).
While on the time scales simulated, the variance for the $\alpha=0.2$ curve perfectly follows the
form $\sigma_M^2(t)=\mbox{const}+D_2t^2$ (a least-square fit to this function is 
displayed by a thin solid line), in the case of $\alpha=0.4$, we observe that the data 
do not follow a power law $\sigma_M^2(t)=\mbox{const}+D_{\alpha} t^{\alpha}$, which supports the notion of a time-dependent crossover from ballistic
to diffusive dynamics. In fact, the numerical results
yield a variance that clearly increases linearly in time for  $t\gtrsim 30/J$. Note that the transition from the massless to the massive 
phase in this
model is of the Beresinski-Kosterlitz-Thouless type,\cite{haldane82,okamoto92,white96} with an exponentially growing correlations lengths as the critical point 
$\alpha_{\mathrm{crit}}$ is approached from $\alpha>\alpha_{\mathrm{crit}}$. This renders it very difficult to see a sharp transition
in the transport behavior using exact diagonalization or DMRG as $\alpha_{\mathrm{crit}}$ is crossed.

 We mention, though,  that for other values of $\alpha> \alpha_{\mathrm{crit}}\approx 0.241$ (results not shown here), on similar time-scales, no diffusive behavior is seen.
Moreover, the time-scale at which diffusive dynamics emerges seems to strongly depend  on $B_0$, {\it i.e.}, on how far the initial state is perturbed over the actual ground state.
The qualitative trend is that the larger $B_0$, the faster diffusive transport is established. Unfortunately, the larger $B_0$, the worse is the entanglement growth which renders
tDMRG simulations more difficult (see, {\it e.g.}, Ref.~\onlinecite{dechiara06}).

Keeping in mind these remarks,  we are in a position to  conjecture that in general, in massive phases of one-dimensional spin models, spin transport
is diffusive. Combined with the existing results for the high-temperature
limit,\cite{hm03,hm07a,zotos04} our work suggests  that this 
observation applies independently of temperature.

Note that a recent exact diagonalization study\cite{santos08} has promoted a different behavior, namely evidence 
for ballistic spin transport at zero temperature in the frustrated spin chain at $\alpha=1$. This conclusion is based  
on the presence of certain oscillations (dubbed {\it a bouncing behavior})\cite{santos08,steinigeweg06} in $M_i(t)$ in the evolution from an initial state
with all spins pointing up(down) in the left(right) part of an open system (compare
Refs.~\onlinecite{antal99,gobert05}).
We believe that our analysis of the variance is more quantitative, which may imply that the oscillations in $M_i(t)$ reported on in 
Ref.~\onlinecite{santos08} are possibly of a different origin. A recent study of quantum quenches in the $XXZ$ chain proposes that oscillations seen in the
order parameter are related to  the quantum phase transition at $\Delta=1$
(Ref.~\onlinecite{barmettler09}, see also Ref.~\onlinecite{barthel08a}).

\section{Summary and Discussion}
\label{sec:sum}

In this work we studied the nonequilibrium magnetization dynamics in one-dimensional
spin models at zero temperature using the adaptive time-dependent DMRG method on system sizes as large as $L=200$ sites. We considered several models:
the integrable spin-1/2 $XXZ$ chain, the frustrated chain, and the two-leg spin ladder.
Based on the analysis of the time-dependence of the spatial variance of the magnetization during the 
time evolution starting from initial states with an inhomogeneous magnetization profile, we conclude 
that in the critical regime of the $XXZ$ chain, the magnetization dynamics is  ballistic. In contrast to that, 
in the massive regime, our results indicate diffusive transport at half filling, while
ballistic transport is restored away 
from half filling. 
A major aspect of our work is that we scanned the entire regime going from small 
 to very strong perturbations over the ground state. This substantially extends previous
studies of linear-response functions as we clearly enter into a regime with the system driven out of equilibrium.
In the case of the massless
regime of the $XXZ$ chain, ballistic transport is seen for substantially perturbed  initial state,
while for the most extreme initial states, {\it i.e.}, pure Fock states, we still find a power-law for the time-dependence of
the variance, but, on the times scales simulated, with an exponent below two.\cite{gobert05}

As for the nonintegrable models, the frustrated chain and the ladder, the numerical data clearly support the notion of 
diffusive dynamics in the ladder system. In the case of the frustrated chain, our data are consistent with a  transition
from ballistic to diffusive behavior as the quantum critical point $\alpha_{\mathrm{crit}}\approx 0.241$ is 
crossed. In the limit of small perturbations, this result confirms with the general picture that
massless phases support ballistic, and massive ones diffusive dynamics, at zero
temperature and irrespective of integrability.\cite{scalapino92}
 Overall, 
a difference between the low- and high-temperature behavior is then evident: numerical results for the
high-temperature limit\cite{hm07a}
consistently support the notion of vanishing Drude weights in nonintegrable models, and thus normal transport behavior, irrespective of
what the ground state phases are. 
Conversely,  exact diagonalization studies find a
finite Drude weight in the gapless phase of the frustrated spin-1/2 chain at zero temperature.\cite{bonca94,vieira01,furukawa08}
In the low, or more extremely, zero temperature case, effective low-energy theories
are expected to give a valid description, which typically predict diverging transport coefficients
of clean spin systems (see, {\it e.g.}, Ref.~\onlinecite{saito03}). 
As for the heat transport measurements on spin chain and ladders experiments (see
Refs.~\onlinecite{sologubenko07a, hess07a} for a survey), a dominant  magnetic contribution is usually
evident in the {\it high}-temperature regime, where the validity of effective low-energy theories for the description of transport
is not obvious.

Finally, the approach of distinguishing ballistic from diffusive transport by analyzing the spatial variance of a density 
like quantity could be instrumental in characterizing ultracold atomic gases in  optical lattices as well. There, one typically 
realizes the expansion of particles into  an empty lattice, and experimentally it is possible to 
measure the expanding cloud's radius.
It  would thus be very interesting to identify conditions for ballistic as compared to diffusive dynamics for model
systems typically encountered in ultracold atomic gases such as the Bose-Hubbard model or a
two-component Fermi gas in an optical lattice.

\begin{acknowledgments}
We are grateful to W. Brenig, A. Feiguin  and A. Kolezhuk for fruitful discussions.
S.L., F.H.-M., and U.S. acknowledge support from the {\it Deutsche Forschungsgemeinschaft} through FOR 912.
\end{acknowledgments}


\end{document}